\documentclass[aps,pre,showpacs,preprint]{revtex4}
\usepackage{graphicx}
\usepackage{bm}
\usepackage{amsmath,amssymb}

\begin{document}
\title{Elasticity of a polydisperse hard-sphere crystal}

\author{ Mingcheng Yang}
\author{Hongru Ma}
\email{hrma@sjtu.edu.cn}

\affiliation{Institute of Theoretical Physics, Shanghai Jiao Tong
University, Shanghai 200240, People's Republic of China}
\date{\today}

\begin{abstract}

A general Monte Carlo simulation method of calculating the elastic
constants of polydisperse hard-sphere colloidal crystal was
developed. The elastic constants of a size polydisperse hard sphere
$fcc$ crystal is calculated. The pressure and three elastic
 constants($C_{11}$, $C_{12}$ and $C_{44}$) increase significantly with the polydispersity.
It was also found from extrapolation  that there is a mechanical terminal polydispersity
above which a $fcc$ crystal will be mechanically unstable.

\end{abstract}


\maketitle

\section{Introduction}

Elastic constants are among the most important physical quantities
describing macroscopic mechanical behaviors of a crystal. For  hard
sphere crystals, the elastic constants had been calculated by
various groups with different methods including density functional
theory \cite{Jaric,Jones,Velasco,Xu,Laird}, Monte Carlo simulations
\cite{Runge,Farago,Sengupta,Kwak,Tretiakova} and molecular dynamics
simulations \cite{Frenkel,Pronk} in the last two decades.  The work
of Jaric \cite{Jaric} and Jones \cite{Jones} predicted a negative
poisson ratio in the hard sphere crystal, which stimulated much
interest in the investigation of  the elasticity of this system.
However, their predictions were proved incorrect by subsequent
studies \cite{Frenkel, Runge,Velasco,Xu}. The hard sphere system is
the simplest in the systems which have pure repulsion interaction,
thus the study of elasticity of hard sphere crystals is an excellent
starting point  to the studies of more complicated and realistic
repulsive systems. Due to the simplicity of the hard sphere system,
it often serves as a simple model system for testing new theoretical
approaches. Hard sphere model is also an important model for a large
class of colloid systems, understandings of colloid systems are
greatly benefited from the extensive researches of the monodisperse
hard sphere model. A more realistic model describing hard sphere
colloidal systems is the size polydisperse hard sphere system. The
size polydispersity of colloid particles is an intrinsic property of
colloidal systems \cite{Blaaderen}. The polydispersity of colloidal
hard spheres is characterized by the ratio of the standard deviation
to the mean of the diameter. It has a remarkable influence on the
thermodynamic and dynamic behaviors of the system
\cite{Pusey,Bolhuis,Fasolo,Phan,Russel,Martin,
Dullens,Schope,Villeneuve}. It is natural to expect that the elastic
constants of a polydisperse hard sphere crystal differ from those of
a monodisperse hard sphere crystal. However, it seems that the
problem of elastic constants of polydisperse hard sphere crystals is
not properly addressed in the literature to the best of our
knowledge.

In this work we propose a general Monte Carlo scheme to investigate
the elasticity of a polydisperse hard sphere system, by which we
calculated the elastic constants of a size polydisperse hard sphere
crystal. We only consider face-center-cubic ($fcc$) crystal
structure in this study, as our previous calculations \cite{Yang}
showed that $fcc$ structure is still the most stable one for a size
polydisperse hard sphere crystal as in the monodisperse case. When
simulating a polydisperse crystal the semigrand ensemble is the best
candidate \cite{Bolhuis,Kofke, Bates}. In the ensemble the imposed
physical quantity is the chemical potential difference $\Delta \mu$
of particles of each kind to a reference kind. To obtain the
chemical potential difference for a prescribed size distribution
$\rho(\sigma)$, one has to solve a functional inverse problem
\cite{Escobedo, Wilding,Wilding1}, which can be accomplished easily
by the semigrand ensemble version of  the \emph{nonequilibrium
potential refinement} (NEPR)  method (SNEPR method) \cite{Yang}.

Simulation approaches for calculating elastic constants of a
monodisperse hard sphere crystal are divided into two categories.
One is the ``fluctuation'' method \cite{Farago} where elastic
constants are related to the thermal averages of the corresponding
stress components. There are some difficulties to extend the method
to the polydisperse system. In the present paper we employ another
method, the so called ``strain'' method \cite{Runge,Sengupta,
Tretiakova,Frenkel,Pronk}, where elastic constants can be obtained
from the free energy-strain relation or its first derivative, the
stress-strain relation. In the simulation we used the extended
ensemble method \cite{Yang,Yang1} to determine the Helmholtz free
energy of the crystal with different strain, which can also be
obtained by thermodynamic integration method \cite{Smit}. Then the
elastic constants were extracted from the free energy-strain data.

The paper is organized as follows. In Sec. II, we introduce the
model and explain how the semigrand ensemble can be applied to
calculate the elastic constants of polydisperse hard sphere
crystals. Section III describes the simulation method employed in
this work. The computational details and results are provided in Sec.
IV. Finally, we present our conclusions in Sec. V.

\section{Model and Theory}

\subsection{Semigrand canonical emsemble}

The semigrand canonical ensemble (SCE) is the most suitable ensemble
for the simulation study of the elastic properties of a size
polydisperse hard sphere  crystal. In this ensemble the total number
of particles and the distribution of particle sizes are fixed while
the number of particles of each size is permitted to fluctuate. In
the simulation study, the total number of particles in the system
usually limited to a few hundreds to thousands, which are too small
to distribute accurately  according to a prescribed distribution of
particle sizes. With the semigrand canonical ensemble, the
distribution is realized through averages since the particle sizes
are allowed to fluctuate. The grand canonical ensemble can also
achieve the goal of distribution realization but insertion and
deletion of particles often require more computational resources.
   For a size
polydisperse system consisting of $N$ particles with the composition
distribution $\rho(\sigma)$ in a constant volume $V$, the Helmholtz
free energy $F(N,V,\{\rho(\sigma)\},T)$ of the system is given by
\begin{equation}
F=-PV+N\int\mu(\sigma)\rho(\sigma)d\sigma,
\end{equation}
where $P$ is the pressure, $\sigma$ is the diameter of the particles,
and $\mu(\sigma)$ is the chemical potential of particles with diameter $\sigma$.
 The semigrand canonical
free energy(SCFE) $Y(N,V,\{\mu(\sigma)\},T)$ is obtained from the Helmholtz free
energy by the Legendre transformation,
\begin{equation}
Y=F-N\int(\mu(\sigma)-\mu(\sigma_{r}))\rho(\sigma)d\sigma,
\end{equation}
where $\sigma_{r}$ is the diameter of an arbitrarily chosen
reference component. The SCFE $Y(N,V,\{\mu(\sigma)\},T)$ is a
functional of $\mu(\sigma)-\mu(\sigma_{r})$. In semigrand canonical
ensemble the thermodynamic variable to characterize the equilibrium
system is $\mu(\sigma)-\mu(\sigma_{r})$ rather than the composition
distribution $\rho(\sigma)$.

The partition function $\gamma$ for this ensemble is
\cite{Briano}
\begin{eqnarray}
\gamma &=&\frac{1}{N!}\int_{\sigma_{1}}\cdots\int_{\sigma_{N}}
Z_{N}\left[\prod_{i=1}^{N}
\frac{1}{\Lambda^{3}(\sigma_{i})}\right]\times
\exp\left\{\beta\sum_{i=1}^{N}(\mu(\sigma_{i})-\mu(\sigma_{r}))\right\}
\prod_{i=1}^{N}d\sigma_{i},
\end{eqnarray}
here $\Lambda(\sigma_{i})=h/(2\pi m_{i}kT)^{1/2}$ is the thermal
wavelength of the \emph{i}th   particle and $Z_{N}$ is the canonical
configuration integral for a given size configuration
\begin{equation}
Z_{N}=\int_{r_{1}}\cdots\int_{r_{N}}e^{-\beta
U}\prod_{i=1}^{N}d\textbf{r}_{i}.
\end{equation}
By introducing the excess chemical potential relative to ideal gas
$\mu_{ex}(\sigma_{i})=\mu(\sigma_{i})
-kT\ln(\frac{N\Lambda(\sigma_{i})^{3}}{V})$, we can  rewrite
$\gamma$ in a more symmetric form
\begin{eqnarray}
\gamma
=\frac{1}{N!\Lambda^{3N}(\sigma_{r})}\int_{\sigma_{1}}\cdots\int_{\sigma_{N}}
Z_{N}\times
\exp\left\{\beta\sum_{i=1}^{N}(\mu_{ex}(\sigma_{i})-\mu_{ex}(\sigma_{r}))\right\}
\prod_{i=1}^{N}d\sigma_{i}.
\end{eqnarray}
The partition function $\gamma$ is related to the thermodynamic
potential $Y$ through
\begin{equation}
Y=-kT\ln\gamma(N,V,T,{\mu_{ex}(\sigma)-\mu_{ex}(\sigma_{r})}).
\label{eq6}
\end{equation}

In practical simulations the diameter of
particles are discretized into a series of special particle sizes or
divide the total size range of particles into many small bins
\cite{Wilding,Wilding1}. As a result, the semigrand canonical
partition function $\gamma$ becomes
\begin{equation}
\gamma=\frac{1}{N!\Lambda^{3N}(\sigma_{r})}\sum_{\sigma_{1}=\sigma_{min}}^{\sigma_{max}}\cdots
\sum_{\sigma_{N}=\sigma_{min}}^{\sigma_{max}} Z_{N}\times
\exp\left\{\beta\sum_{i=1}^{N}(\mu_{ex}(\sigma_{i})-\mu_{ex}(\sigma_{r}))\right\},
\end{equation}
here $\sigma_{min}$ and $\sigma_{max}$ are, respectively, the
minimum and maximum of particle sizes.

\subsection{Elastic constants of size polydisperse solid}

The elasticity of a solid is the property that the solid deforms in
response to an external stress and return to its initial
configuration when the stress is removed. Usually the deformation is
small, otherwise the deformation may become permanent. In the case
of hard sphere solids, the situation is more complicated because an
external stress is necessary to stabilize the solid, while the
deformation of solid is induced by exerting an excess external
stress. The deformation of a continuous solid can be described by
the Lagrangian strain tensor
\begin{equation}
\eta_{ij}=\frac{1}{2}\left(\frac{\partial u_{i}}{\partial{x_{j}}}
+\frac{\partial u_{j}}{\partial{x_{i}}}+\frac{\partial
u_{k}}{\partial{x_{i}}}\frac{\partial
u_{k}}{\partial{x_{j}}}\right),
\end{equation}
here $u_{i}$ is the $i$th component of the displacement field, and
$x_{i}$ is the $i$th component of the position of the displaced
point in the solid, repeated indices are summed from $1$ to $3$. The
Helmholtz free energy density $f=F/V$(where $V$ is the volume of the
undeformed solid) is a functional of the strain field. For a
homogeneous deformation, namely the Lagrangian strain tensor
$\eta_{ij}$ is independent of the position, the free energy density
becomes a function of the constant strain. The elastic constants can
be defined in terms of the following Taylor expansion of the free
energy density
\begin{equation}
f(\eta_{ij})=f(\textbf{0})+T_{ij}(\textbf{0})\eta_{ij}+\frac{1}{2}C_{ijkl}
\eta_{ij}\eta_{kl}+\cdots,\label{eq9}
\end{equation}
where $f(\textbf{0})$ is the Helmholtz free energy density of the
unstrained solid, and $T_{ij}(\textbf{0})$ is the stress tensor of
the unstrained solid which is necessary to stabilize the hard sphere
crystal, in the case of an $fcc$ hard sphere solid,
$T_{ij}(\textbf{0})=-p\delta_{ij}$ where $p$ is the hydrostatic
pressure. The $C_{ijkl}$ is the second-order elastic constants.

It should be emphasized that in the case of a size polydisperse
system, the composition distribution of the strained system must
remain the same as that of the unstrained system, this is because
$\eta_{ij}$ and $\rho(\sigma)$ are two independent variables of the
Helmholtz free energy. Thus the explicit expression of elastic
constants is
\begin{equation}
C_{ijkl}=\frac{\partial^{2}f(\eta )}{\partial\eta_{ij}\partial
\eta_{kl}}\bigg|_{\rho(\sigma);\eta=0}.
\end{equation}
Here the subscripts $\rho(\sigma)$ means that the distribution of particle sizes
is fixed during the application of strain. In semigrand canonical ensemble, the Helmholtz
free energy density
of the strained system with strain tensor $\eta$ and composition
distribution $\rho(\sigma)$ is
\begin{eqnarray}
f(\eta)&=&y(\eta)+\frac{N}{V}\int
(\mu(\sigma,\eta)-\mu(\sigma_{r},\eta) ) \rho(\sigma)d\sigma
\nonumber\\
&=&y(\eta)+\frac{N}{V}\int(\mu_{ex}(\sigma,\eta)-\mu_{ex}(\sigma_{r},\eta))
\rho(\sigma)d\sigma +\frac{3NkT}{V}\int \ln\left(\frac{ \Lambda
(\sigma) }{\Lambda(\sigma_{r})}\right)  \rho(\sigma)d\sigma,
                             \label{eq11}
\end{eqnarray}
here $y=Y/V$. In the semigrand ensemble the composition distribution
$\rho(\sigma)$   depends not only on the chemical potential
difference but also on the strain $\eta_{ij}$. Therefore, in
order to keep the composition distribution unchanged, the chemical
potential difference
$\mu_{ex}(\sigma,\eta)-\mu_{ex}(\sigma_{r},\eta)$  has to be
adjusted for each strain $\eta_{ij}$. That means at each given
strain the chemical potential difference is recalculated. The
evaluation of the chemical potential difference for a given
composition distribution and strain can easily be performed by the
SNEPR method \cite{Yang}, explained in the following subsection.

For the size polydisperse hard sphere $fcc$ crystal, there are only
three independent second-order elastic constants $C_{1111}$,
$C_{1122}$ and $C_{1212}$. In the Voigt notation they can be written
in  a more compact format $C_{11}=C_{1111}$, $C_{12}=C_{1122}$ and
$C_{44}=C_{1212}$. It should be noted that the elastic constants
defined in this way is not the one that is measured directly in
experiments, though it is widely used in theoretical
investigations\cite{Frenkel,Runge,Laird,Pronk,Kwak,Tretiakova}.

For a cubic crystal(including $fcc$ structure) under isotropic
pressure $P$ the experimentally measured elastic constants, $C^{T}$,  are related to the above
defined $C$ by the following
relations \cite{Xu}
\begin{equation}
C^T_{11}=C_{11}-P; ~~~~C^T_{12}=C_{12}+P; ~~~~C^T_{44}=C_{44}-P.
\end{equation}
It is only when $P=0$ that the two sets of elastic constants
coincide. A detail discussion of the elastic constants under stress
can be found in \cite{Barron}.

\section{Simulation Method}

\subsection{SNEPR method}

In order to fix the composition distribution of the polydisperse
crystal, we extend the NEPR algorithm \cite{Wilding1} to the
semigrand canonical ensemble. The algorithm can be used to find the
chemical potential difference for a prescribed composition
distribution, i.e.
$\Delta\mu_{ex}(\sigma)=\Delta\mu_{ex}(\{\rho(\sigma)\})$ at a given
strain. Here, we only give a brief description of the method. A
detailed presentation can be found in references
\cite{Yang,Wilding1}. For a given particle size distribution and
strain, the chemical potential difference can be calculated  by a
Monte Carlo iteration procedure. Firstly, initial guess of the
excess chemical potential is assigned to the implement of a
semigrand canonical ensemble simulation, and then it is modified at
every few MC steps according to the instant size distribution
$P_{ins}(\sigma)$ as follows,
\begin{equation}
\Delta\mu^{'}_{ex}(\sigma)=\Delta\mu_{ex}(\sigma)-
\gamma_{i}\left(\frac{P_{ins}(\sigma)-P(\sigma)}{P_{ins}(\sigma)}\right)\qquad
\forall \sigma.
\end{equation}
Here $\gamma_{i}$ is a modification factor of the $i$th iteration.
When the difference of the average size distribution
$\overline{P}(\sigma)$ and the given composition is less than a
specified value $\xi$
 \begin{equation}
\xi\geq
\max\left(\left|\frac{\overline{P}(\sigma)-P(\sigma)}{P(\sigma)}\right|\right),
\end{equation}
one loop of the iteration is finished. The modification factor is
then reduced, and the excess chemical potential of the last
iteration is used as the initial input and start the next iteration.
The iteration continues till the modification factor $\gamma$
reaches a very small value, typically $10^{-5}$,  and the resulted
excess chemical potential is then regarded as the solution of the
problem.

\begin{figure}
\includegraphics[angle=0,width=0.48\textwidth]{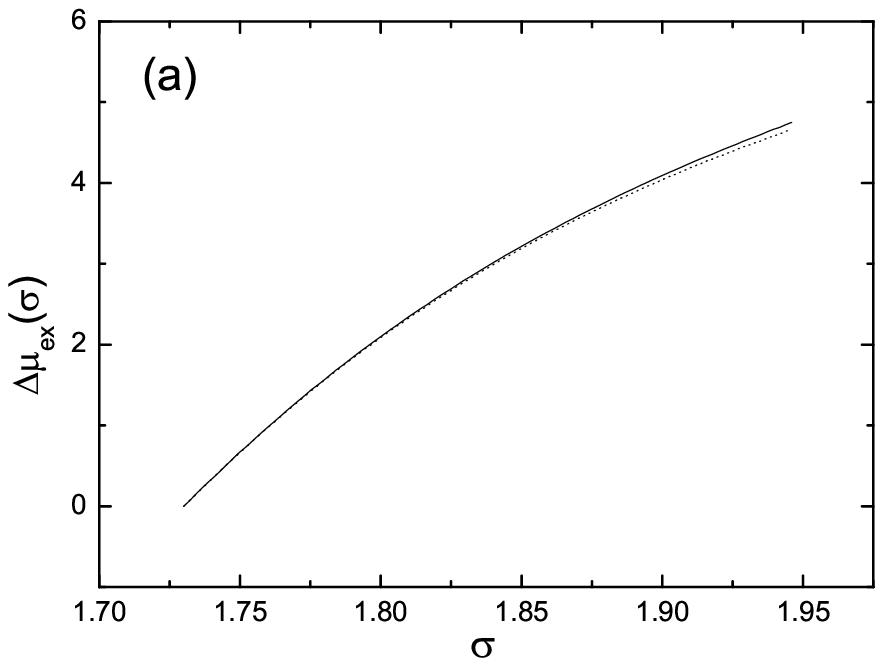}
\includegraphics[angle=0,width=0.5\textwidth]{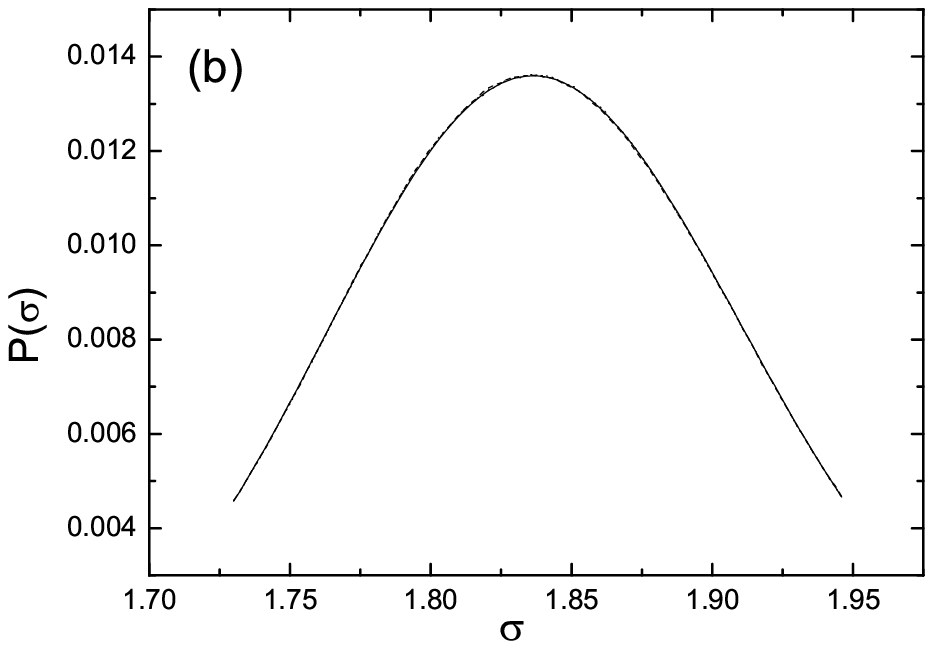}
\caption{(a), the solved excess chemical potential as function of
the diameter of particles. Dotted line corresponds to the unstrained
crystal, and solid line to the crystal with a contraction strain.
(b), the solid line is the plot of the truncated Schultz function.
The dotted line and dashed line are, respectively, the composition
distribution for the unstrained and strained crystal, which are
obtained from simulation by using the $\Delta\mu_{ex}(\sigma)$
plotted in (a). It is difficult to distinguish the three
distributions.}\label{figsnepr}
\end{figure}

In order to check the validity of SNEPR method, we apply it to both
unstrained and strained polydisperse hard sphere crystals. Figure
1(a) are the calculated excess chemical potential for the truncated
Schultz distribution as function of the diameter of particles. The
lower dotted line and upper solid line correspond to the unstrained
crystal and the crystal with a contraction strain $0.004$,
respectively. In the contracted crystal it is more difficult to
increase the particle size, so a higher excess chemical potential is
necessary to fix the composition distribution. Figure 1(b) is the
comparison of the given Schultz distribution and the distribution
generated with the calculated excess chemical potential. In figure
1(b) we plotted three lines of the distribution, the solid line is
the given distribution, dotted line is the distribution generated
from the chemical potential difference for the unstrained crystal,
and dashed line is the one for the strained crystal. The agreement
among the three distributions is  excellent and nearly
undistinguishable from the figure.

\subsection{free energy difference calculation}

To determine the elastic constants one only needs to know the Helmholtz free
energy difference between the unstrained system and the strained system with
the same composition distribution. The Helmholtz free energy difference can be
decomposed into two parts according to equation (\ref{eq11})
\begin{equation}
\Delta  f(\eta)=\Delta y(\eta) +\Delta
\left\{\frac{N}{V}\int(\mu_{ex}(\sigma,\eta)-\mu_{ex}(\sigma_{r},\eta))
\rho(\sigma)d\sigma\right\},  \label{eq12}
\end{equation}
where $\Delta$ denotes the excess quantities relative to the
unstrained system. Correspondingly, the elastic constants becomes
\begin{equation}
C_{ijkl}=\frac{\partial^{2}[\Delta f(\eta)]}{\partial\eta_{ij}\partial
\eta_{kl}}\bigg|_{\rho(\sigma);\eta=0}.
\end{equation}

The second term on the right hand side of  (\ref{eq12}) can be
calculated when the
$\mu_{ex}(\sigma_{i},\eta)-\mu_{ex}(\sigma_{r},\eta)$ for the fixed
composition is determined. Our main task is thus to compute the
difference of semigrand free energy $\Delta y(\eta)$. To this end an
extended ensemble \cite{Lyubartsev} is introduced. The Lagrangian
strain tensor $\eta$ is regarded as an additional ensemble variable
and different $\eta$ corresponds to different macroscopic state. The
partition function of the extended ensemble is defined as
\begin{equation}
\Gamma(N,T,\rho(\sigma))=\sum_{\eta=0}^{\eta_{max}}\gamma(N,\eta,T,{\mu_{ex}(\sigma,\eta)-\mu_{ex}(\sigma_{r},\eta)}),
\label{eq14}
\end{equation}
where $\eta_{max}$ denotes the state with maximum strain  and all
macroscopic states $\eta$ possess the same composition distribution
$\rho(\sigma)$. From equation (\ref{eq6}) and (\ref{eq14}) the
$\Delta y(\eta)$ becomes
\begin{eqnarray}
\Delta y(\eta) &=&
\ln\left[\frac{\gamma(N,0,T,{\mu_{ex}(\sigma,0)-\mu_{ex}(\sigma_{r},0)})}
{\gamma(N,\eta,T,{\mu_{ex}(\sigma,\eta)-\mu_{ex}(\sigma_{r},\eta)})}\right]
\nonumber\\
&=& \ln\left[\frac{\gamma(0)/\Gamma} {\gamma(\eta)/\Gamma}\right]
\nonumber\\
&=& \ln\left[\frac{Pr(0)}{Pr(\eta)}\right],  \label{eq15}
\end{eqnarray}
where $Pr(\eta)$ is the probability that the system is in the
macroscopic state $\eta$. Therefore, knowledge of the macroscopic
state probability distribution is sufficient to evaluate the elastic
constants. The probability can be calculated from simulation by the
flat histogram methods \cite{Berg,F.Wang,S.Wang}. Here we
demonstrate the implementation in the Wang-Landau scheme, other
schemes can also be implemented.

The extended ensemble Monte Carlo simulation involves three kind of
moves, the first is the particle displacement, the second is the
particle resizing and the third is the deformation of the simulation
box which corresponding to the walk in the stain $\eta$ space. The
first two  moves are accepted or rejected in the usual Metropolis
way, i.e. if no overlap between particles happens, the trial move
$(r,\sigma)\rightarrow (r{'},\sigma{'})$ is accepted with
probability
\begin{equation}
P_{acc}(r,\sigma\rightarrow
r{'},\sigma{'})=\min\{1,\exp(\beta\Delta\mu_{ex}(\sigma{'},\eta)
-\beta\Delta\mu_{ex}(\sigma,\eta))\}.
\end{equation}
Where $\Delta\mu_{ex}(\sigma,\eta)=\mu_{ex}(\sigma,\eta)-\mu_{ex}(\sigma_r,\eta)$.
The trial move in the $\eta$ space is treated with the Wang-Landau sampling in order to
obtain the macroscopic state probability distribution $Pr(\eta)$. With a initial guess of the
$Pr(\eta)$, the acceptance/rejection criteria for the simulation box deformation, $\eta \rightarrow \eta'$
 is
\begin{equation}
P_{acc}(\eta\rightarrow\eta{'})= \left\{
\begin{array} {ll}
\min\left\{1,\frac{Pr(\eta)}{Pr(\eta{'})}\left(\frac{V{'}}{V}\right)^{N}
e^ {\beta\sum_{\sigma}(\Delta\mu_{ex}(\sigma,\eta{'})
-\Delta\mu_{ex}(\sigma,\eta))}\right\} \quad & \textrm{if no overlap of spheres}
\\0 & \textrm{otherwise}
\end{array},
\right.
\end{equation}
where $\eta$ only takes some discrete values $\eta_{1}$,
$\eta_{2}$...$\eta_{n}$. The chemical potential difference for each strain $\eta$
were calculated with the SNEPR and stored before the simulation with
extended ensemble.  During the extended ensemble simulation the (unnormalized) $Pr(\eta)$ is
updated by multiplying a modification factor $f>1$ when a state of $\eta$ is visited,
a histogram of the distribution in $\eta$ of the visited states is recorded to monitor
the convergence of $Pr(\eta)$. The relative probability distribution of $\eta$ is
obtained at the end of the simulation, which is then used  to determine
$\Delta y(\eta)$ from equation (\ref{eq15}). Because the
composition distribution $\rho(\sigma)$ is prescribed in advance,
substituting $\rho(\sigma)$, $\Delta\mu_{ex}(\sigma,\eta)$ and
$\Delta y(\eta)$ into (\ref{eq12}) the Helmholtz free energy
density difference $\Delta f(\eta)$ can then be determined.
Finally, the elastic constants can be obtained from a polynomial fit
to the free energy-strain data.

\section{Simulation details and Results}
Most of the simulations are performed with a system of $256$ size polydisperse hard spheres
in a parallelepiped box, periodic boundary conditions are used in all three directions.
 The initial configuration is an
ideal $fcc$ crystal(zero strain). The density of undeformed crystal
taken in this simulation is $\rho=0.576$. These value is chosen
because there are some known results of the terminal polydispersity
\cite{Bolhuis} and the elastic constants in the monodisperse case
\cite{Pronk} in literature. As the previous computations
\cite{Frenkel,Runge,Kwak,Tretiakova} the elastic constants of $fcc$
crystal can be determined completely from three independent
deformations. They are the contraction, contraction-expansion and
shear deformation. The composition distribution of particles used in
the simulation is the truncated Schultz function
\begin{equation}
P(\sigma)=\frac{1}{z!}\left(\frac{z+1}{\overline{\sigma}}\right)^{z+1}\sigma^{z}
\exp\left[-\left(\frac{z+1}{\overline\sigma}\right)\sigma\right]
\qquad \sigma_{min}\leq\sigma\leq\sigma_{max},
\end{equation}
where $\sigma_{min}$ and $\sigma_{max}$ are respectively the minimum
and the maximum diameters of hard sphere particles.
$\overline\sigma$ is the average diameter and $z=\delta^{-2}-1$
controls the width of the distribution. The criteria of the
truncation is that the probability density at both ends of the
distribution are almost equal. They are several times less than the
maximum probability density, as shown in figure \ref{figsnepr}(b).
Here the probability density at $\sigma_{min}$ and $\sigma_{max}$ is
$1/3$ of the peak value.  In the present paper the effect of the
cutoff is not studied, the emphasis is on the effect of
polydispersity to the elastic properties. The size polydisperse
degree is defined by
$\delta=\frac{\sqrt{\overline{(\sigma-\overline{\sigma})^{2}}}}
{\overline{\sigma}}$. In the simulation we consider a uniform
discrete set of diameters of particles. When the number of discrete
diameters is large enough, the diameter of particles tends to a
continuous variable which can resemble the real polydisperse system.
In this study $101$ different sizes of diameters were used.

As mentioned above, the composition depends not only on the chemical
potential difference but also on the strain tensor. Therefore, the
chemical potential differences are calculated before the extended
ensemble simulation is performed. The results indicate that the
chemical potential difference increases with the magnitude of the
strain, as plotted in Figure \ref{figsnepr}(a). In the case of a
contraction deformation, the increase of the chemical potential
difference is more significant than the case of shear deformation.
This is reasonable because the contraction deformation consumes more
configuration space so that the turn to larger sizes is more
difficult and requires a larger chemical potential.

\begin{table}[!h]
\tabcolsep 0pt \caption{Elastic constants of $fcc$ hard sphere
crystals and simulation parameters. All undeformed systems have the
same density $\rho=0.576$. Here, $\delta$ is the size plydispersity,
$N$ the number of particles and $P$ the pressure of the hard sphere
crystal. The first two rows are the elastic
constants of monodisperse hard sphere crystals obtained from
reference \cite{Pronk}. }                    \label{table1}
 \vspace*{-12pt}
\begin{center}
\def\temptablewidth{0.6\textwidth}
{\rule{\temptablewidth}{2pt}}
\begin{tabular*}
{\temptablewidth}{@{\extracolsep{\fill}} c|c|c|c|c|c}
$\delta$ & $N$ & $P$      & $C_{11}$  &$C_{12}$ &$C_{44}$  \\
\hline
    0    &256  &          & 115.5(10) &32.0(4)  &72.4(3.2) \\
    0    &13292&          & 117.4(4.4)&31.54(15)&71.96(11) \\
    0    &256  &14.54(2.5)& 115.1(46) &32.7(16) &72.0(35)  \\
    0.03 &256  &15.49(2)  & 125.1(47) &43.7(14) &77.1(45)  \\
    0.03 &2048 &15.528(6) & 129.7(34) &47.3(12) &76.5(21)  \\
    0.039&256  &16.24(2)  & 135.5(33) &55.9(10) &78.5(45)  \\
    0.05 &256  &17.34(1)  &150.0(28)  &73.6(8)  &83.5(29)

\end{tabular*}
{\rule{\temptablewidth}{1pt}}
\end{center}
\end{table}

\begin{figure}
\centering
\includegraphics[angle=0,width=0.55\textwidth]{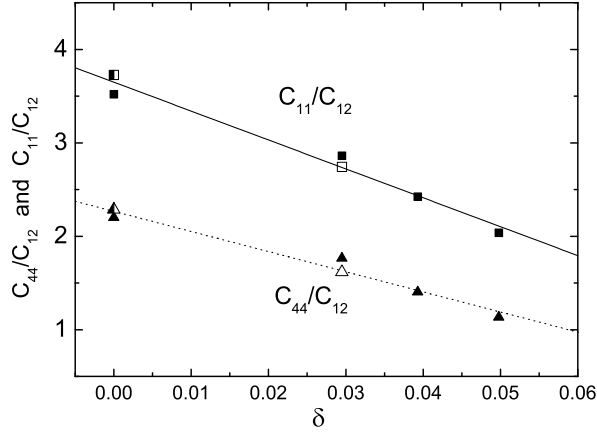}
\caption{The ratio of two elastic constants as a function of the
size polydispersity $\delta$. The triangles and squares indicate
$C_{11}/C_{12}$ and $C_{44}/C_{12}$, respectively. The filled
symbols denote the data obtained from 256-particle system, open
symbols from 2048-particle system. The semi-filled symbols are from a 13292-particle
system given by reference \cite{Pronk}. The  lines
are linear fits to the data.}\label{fig1}
\end{figure}

We performed simulations for four different size polydispersity,
simulation parameters and results are given in Table \ref{table1}.
The maximum polydispersity taken in our simulation is 5\%, because
at higher $\delta$ the crystal may be unstable
\cite{Pusey,Bolhuis,Chaudhuri}. In order to check the system size
dependence we performed simulations for a larger system with the
particle number $N=2048$. The difference between the two systems are
clearly less than or on the same order of the statistical errors,
which indicates that a system of $256$ particles is already large
enough to get reasonable results, similar to the monodisperse hard
sphere system \cite{Pronk, Tretiakova}. We also calculated the
elastic constants of a  monodisperse hard sphere crystal using the
present method, which are in full agreement with the results of
Pronk and Frenkel \cite{Pronk}. From  Table \ref{table1} we see that
the pressure increases with the increasing of the size
polydispersity and the pressure with $\delta=0.05$ is about $20$
percent higher than the monodisperse system, which is consistent
with the previous reports \cite{Phan,Russel} that the size
polydispersity can induce a higher osmotic pressure in the hard
sphere colloidal crystal. A new phenomenon which has not been
reported previously is that as the size polydispersity increases all
elastic constants($C_{11}$, $C_{12}$ and $C_{44}$) of a $fcc$ hard
sphere crystal increase significantly. At $\delta=0.05$, $C_{12}$ is
$1.3$ times larger than that of the monodisperse system. Figure
\ref{fig1} shows the ratios of $C_{11}$ to $C_{12}$ and $C_{44}$ to
$C_{12}$ as a function of the polydispersity $\delta$. It is
interesting to note that for the polydispersity used in the
simulation these ratios nicely follow a linear relation with the
polydispersity. Experimentally, Phan et al have measured the
high-frequency shear modulus for the hard sphere colloidal crystals
\cite{Phan2}. Their results are comparable to the static shear
moduli of the $fcc$ crystal obtained in this study. Furthermore, our
results indicate the static shear modulus also increases with the
polydispersity. We think that the effect can be detected
experimentally from a precise measurement.

The monodisperse hard sphere system was regarded as a representative
model in the description of many different aspects of hard sphere
colloids. We expected that the polydispersity may give only a small
correction to the monodisperse case. The large increase of the
elastic constants with polydispersity indicates that the
monodisperse model has its limitations in describing the real hard
sphere colloids, especially in the elastic properties.  The physics
behind this large increase is still not clear, one general
explanation is that the elastic constants are the second derivatives
of the free energy, which should be more sensitive to the
polydispersity than the free energy itself. A comprehensive
understanding of this enhancement of elastic constants from
polydispersity requires more extensive research which is beyond the
scope of this paper.

\begin{figure}
\centering
\includegraphics[angle=0,width=0.55\textwidth]{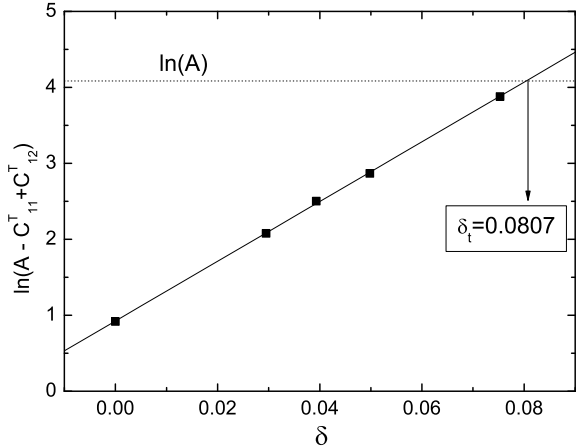}
\caption{The change of $C^T_{11}-C^T_{12}$ with the size
polydispersity $\delta$. The solid line is a linear fit to the data.
The horizontal dotted line represents the function
$y=\texttt{ln(A)}$ and $A$ is a fitted constant(see text). At $\delta=0.0807$ the solid
line intersects with the dotted line. }\label{fig2}
\end{figure}

As is well known, the necessary requirements for the stability of a
cubic crystal(including $fcc$ structure) are
\begin{equation}
C^T_{11}>0; ~~~~C^T_{44}>0; ~~~~C^T_{11}+2C^T_{12}>0;
~~~~(C^T_{11})^2-(C^T_{12})^2>0.
\end{equation}
For the polydisperse hard sphere crystal under consideration, the
first three conditions are obviously satisfied, the last one may be
violated with increasing the polydispersity. Figure \ref{fig2}
depicts the change of $C^T_{11}-C^T_{12}$ with the polydispersity.
The value of $C^T_{11}-C^T_{12}$ decreases with increasing
polydispersity, and tends to zero at $\delta=0.0807$. To get this
terminal polydispersity, we note that the simulation data follows a
relation
\[
C^T_{11}-C^T_{12} =A -e^{\alpha \delta +B},
\]
that means that  $\ln\left(A-C^T_{11}+C^T_{12}\right)$ depends
linearly on the polydispersity $\delta$, as shown in figure
\ref{fig2}. Using the fitted value of $\ln A$ we can obtain the
terminal polydispersity from extrapolation. This defines a
mechanical terminal polydispersity (MTP) where the $fcc$ crystal
becomes unstable. The strain related to the coefficient
$C^T_{11}-C^T_{12}$ is the following contraction-expansion
deformation:
\begin{equation}
x{'}=(1+\epsilon)x, ~~~~y{'}=\frac{1}{1+\epsilon}y, ~~~~z{'}=z.
\end{equation}
That is to say, for $\delta\geq0.0807$ the $fcc$ crystal is no
longer stable under the deformation. The instability can also be
described from the point of view of soft-mode. By solving the
dispersion equation, one easily finds that there exists a transverse
wave, propagates along the diagonal of the $(001)$ crystal plane and
polarized in $xy$-plane, has the dispersion relation
$\rho_m\omega^2=\frac{1}{2}(C^T_{11}-C^T_{12})k^2$ \cite{Landau},
here $\rho_m$ is the mass density, $\omega$ is the circular
frequency and $k$ the wave vector. Therefore, its frequency
decreases substantially as the MTP is approached and this branch of
the wave corresponds to a soft acoustical mode.

For the hard sphere crystal, it is known that there is another
terminal size-polydispersity \cite{Pusey,Bolhuis,Fasolo,Chaudhuri}
above which the crystal will not be the most stable structure, and
disorder solid phase \cite{Chaudhuri} or solid-solid coexistence
phase \cite{Fasolo} may become the most stable equilibrium state. We
refer to the terminal polydispersity as thermodynamical terminal
polydispersity(TTP). The MTP has to be not lower than the TTP, since
above TTP the $fcc$ crystal exists in a metastable state, and has
good mechanical behaviors. Therefore, the MTP obtained in this study
gives an upper limit of the TTP.

One important point of the semigrand canonical ensemble simulation
of polydisperse hard sphere crystal is the swapping of particles. In
this ensemble the particle sizes are allowed to fluctuate to get the
required size distribution, with the fluctuation the particles of
different sizes effectively exchange their spatial positions
constantly, and with sufficient long simulation, the equilibrium
state is realized during the simulation. The calculated elastic
constants in this simulation is the equilibrium elastic constants,
we may refer to them as ideal elastic constants. On the other hand,
in real colloid crystals the particles can not exchange their
positions simply because the free energy barrier is too high to be
overcome in any reasonable time period. The particle arrangements
are basically fixed by the process of crystal growth, which is not
necessarily the equilibrium arrangement. It also noted that during
the strain in the measurement of elastic constants the particles can
only undergo small displacements, and it is not possible for
particles to swap their positions. Based on this observation, a
natural question is that wether the simulated elastic constants is
the same one in a measurement. The answer to this question is
probably yes. The reason is that measurements are always performed
with a crystal of macroscopic size, which contains much more colloid
particles than the number of particles in the simulation study. If
the sample is well relaxed, the self-averaging effect of macroscopic
number of particles may compensate the effect of particle non-swap.
We expect that the ideal elastic constant can be experimentally
measured after sufficient equilibration. To test the possible
deviation from the ideal elastic constants, we also performed
simulations in the canonical ensemble. We randomly picked up several
configurations from the particle sizes distribution, and for each
configuration the particle size then fixed in the subsequent
simulations of applying strains and extracting elastic constants.
For a system of $2048$ particles we find that the elastic constants
with different configuration realizations of the same distribution
give elastic constants within 10\% of difference. The average of the
results from different realization of the configuration is listed in
table \ref{table2}.

\begin{table}[!h]
\tabcolsep 0pt \caption{Comparison of the pressure and elastic
constants obtained from the semigrand canonical ensemble
simulation(upper row) and the canonical ensemble simulation(lower
row).} \label{table2} \vspace*{-10pt}
\begin{center}
\def\temptablewidth{0.6\textwidth}
{\rule{\temptablewidth}{2pt}}
\begin{tabular*}
{\temptablewidth}{@{\extracolsep{\fill}} ccccc}
         & $P$   & $C_{11}$  &$C_{12}$ &$C_{44}$  \\
\hline
   semigrand &15.528& 129.7 &47.3  &76.5 \\
   canonical &15.591   & 137.4 &49.9  &$75.7$

\end{tabular*}
{\rule{\temptablewidth}{1pt}}
\end{center}
\end{table}

\section{Conclusion}

To conclude, the elastic
constants of a $fcc$ polydisperse hard sphere crystal are simulated by the Monte Carlo
method with a semigrand ensemble, the composition distribution is fixed in the simulation
by the SNEPR method. The results show that both the pressure of the hard sphere solid and
the three elastic constants increase with the size polydispersity $\delta$. 
The method can be extended to the soft sphere system and
the system with other polydisperse attributes in a straightforward
manner. Our results also indicate that there is a MTP where the $fcc$
crystal is unstable, which provides us an upper limit of TTP.
Above TTP we do not know which of the two structure(disorder solid
and solid-solid coexistence phase) is the most stable one. Determining
the stable state from computer simulations requires more effort to accomplish.

\begin{acknowledgments}

The work is supported by the National Natural Science Foundation of
China under grant  No.10334020  and in part by the National Minister
of Education Program for Changjiang Scholars and Innovative Research
Team in University. We thank the referee of the original manuscript to point out
the possible difference when particles are not allowed to swap which is discussed
in this revised one.

\end{acknowledgments}

\end{document}